\documentclass{aastex}

\newcommand{\dsh}{(D/H)}

\slugcomment{FUSE ApJL draft}

\usepackage{emulateapj5}

\begin{document}


\title{FUSE observations of the HD molecule toward HD~73882}



\author{R.~Ferlet\altaffilmark{1}, M.~Andr\'e\altaffilmark{2}, 
G.~H\'ebrard\altaffilmark{1}, A.~Lecavelier des Etangs\altaffilmark{1}, 
M.~Lemoine\altaffilmark{3}, G.~Pineau des For\^ets\altaffilmark{4},
E.~Roueff\altaffilmark{4}, B.L. Rachford\altaffilmark{5},
J.M. Shull\altaffilmark{5}, T.P. Snow\altaffilmark{5},
J.B. Tumlinson\altaffilmark{5}, A.~Vidal-Madjar\altaffilmark{1},
D.G.~York\altaffilmark{6}, and H.W. Moos\altaffilmark{2}}

\altaffiltext{1}{Institut d'Astrophysique de Paris, CNRS, 98 bis bld Arago,
F-75014 Paris, France}
\altaffiltext{2}{Department of Physics and Astronomy, Johns Hopkins University, 
Baltimore, MD 21218, USA}
\altaffiltext{3}{DARC, UMR--8629 CNRS, Observatoire de Paris-Meudon, 
F-92195 Meudon, France}
\altaffiltext{4}{DAEC, Observatoire de Paris, F-92195 Meudon, France}
\altaffiltext{5}{CASA, University of Colorado, Boulder, CO 80309, USA}
\altaffiltext{6}{University of Chicago, Department of Astronomy and Astrophysics,
Chicago, IL 6063, USA}


\begin{abstract}
The Lyman and Werner band systems of deuterated molecular hydrogen (HD) occur 
in the far UV range below 1200 \AA. The high sensitivity of the {\it FUSE} 
mission can give access, at moderate resolution, to hot stars shining 
through translucent clouds, in the hope of observing molecular cores in which
deuterium is essentially in the form of HD. Thus, the measurement of the
HD/H$_{2}$ ratio may become a new powerful tool to evaluate the deuterium
abundance, D/H, in the interstellar medium. We report here on the detection of 
HD toward the high extinction star HD 73882 ($E_{\rm B-V}=0.72$). A 
preliminary analysis is presented.
\end{abstract}




\section{Introduction.}

It has long been recognized that the primordial abundance of deuterium
represents the most sensitive probe of the baryonic density
$\Omega_{\rm b}$~of the Universe (see, e.g. Schramm \& Turner 1998;
Olive, Steigman \& Walker 1999). Moreover, the abundance of deuterium at any 
epoch is a lower limit to its primordial abundance since deuterium is solely
destroyed in stars of any mass. Therefore, D/H is also an efficient tracer of 
the universal star formation rate. However, the evolution of the deuterium 
abundance from zero to solar metallicity is still unclear.

Measurements of the atomic \dsh~ratio have been performed in different
astrophysical sites, namely in moderate to high redshift quasar absorbers, 
in the presolar nebula and in the local interstellar medium (for reviews, see
e.g. Ferlet \& Lemoine 1996; Linsky 1998; Vidal-Madjar et al. 1998a; Lemoine 
et al. 1999). These studies indicate that D/H may vary within the local
interstellar medium by a factor as high as $\sim2$~over spatial scales of few
tens of parsecs (Vidal-Madjar et al. 1998b; Jenkins et al. 1999; Sonneborn
et al. 2000), while presolar and quasar absorbers D/H abundances are limited 
by the existing scatter in the results.

Deuterated molecules are another means of estimating the deuterium abundance. 
To date, over 20 single D-bearing species and two doubly deuterated molecules, 
D$_2$CO and ND$_2$H, have been observed at radio frequencies both in cold 
interstellar dark clouds and in warmer star forming regions (see e.g. Roueff 
et al. 2000). However, chemical fractionation takes place in cold regions and 
mantle desorption of grains are often invoked in star forming regions. 
Deriving accurate deuterium fractional abundances using these molecules is 
therefore very difficult. 

Recently, the R(2) transition at $37.7\,\mu$m of HD has been detected with 
{\it ISO} in giant planets (Feuchtgruber et al. 1999) and the pure rotational 
J=1$\rightarrow$0 line at $112\,\mu$m toward the Orion Bar (Wright et al. 
1999). Bertoldi et al. (1999) have also detected the excited rotational line 
at $19.43\,\mu$m of HD J=6$\rightarrow$5 in Orion KL, a molecular outflow 
region. Although {\it ISO} thus opened the sky to HD emission, the derived 
column densities depend strongly on the modelling of HD excitation and on 
extinction corrections. With {\it Copernicus}, H$_2$ and HD molecules were 
observed in absorption in the ultraviolet in diffuse interstellar clouds such 
as that toward $\zeta$~Oph (Wright \& Morton 1979). However, the low 
HD/H$_2$~value found (few$\times10^{-7}$~to few$\times10^{-6}$) reflects the 
mostly atomic nature of these diffuse clouds; to determine the D/H ratio from 
these data requires a detailed model of the formation and of the destruction 
of HD.

Because of its high throughput, the Far Ultraviolet Spectroscopic Explorer 
({\it FUSE}; Moos et al. 2000) can give access to denser molecular clouds 
within which one might expect H and D to be essentially in their molecular 
form. If so, no chemical fractionation correction will be required and 
accurate deuterium abundances could result directly from the measurement of 
the HD/H$_2$~ratio. In Section 2, we present {\it FUSE} observations of HD in 
the translucent cloud in front of the star HD 73882, along with the data 
reduction and analysis. A companion paper by Snow et al. (2000) describes the 
H2 observations. A preliminary discussion is given in Section 3, while some 
conclusions are drawn in Section 4.

\section{{\it FUSE} Observations of HD 73882.}

Twenty-seven years after {\it Copernicus}, {\it FUSE} was successfully 
launched on June 24, 1999 from Cape Canaveral. It has a sensitivity of about 
$10^4$~times that of {\it Copernicus}, in the wavelength range from slightly 
below the Lyman limit (905 \AA) to 1187 \AA~(Moos et al. 2000). The higher
throughput comes about because of the use of a guider that can be used for
fainter stars and of the use of an array detector instead of a scanning 
photomultiplier tube. Many absorption lines from different rotational states 
of H$_2$~and HD are available within that range.

The bright early type (O8.5V) star HD 73882 at about 925 pc has been the first
target shining through a translucent cloud, with a reddening 
$E_{\rm B-V}$=0.72 or A$_{\rm v}$=2.44 (note that $E_{\rm B-V}$=0.38 and 
A$_{\rm v}\sim$1 were the highest values observable by {\it Copernicus}), to 
be observed with {\it FUSE} in the time-tagged mode, for a total integration 
time of $\simeq 7\,$hours between 30 and 31 October 1999. At this time, the 
spectrograph was still unfocused, resulting in a spectral resolving power 
R=$\lambda/\Delta\lambda\simeq14000$. Unfortunately, after a short time the 
two short-wavelength SiC channels became misaligned and failed to accumulate 
on the target. However, for our very reddened target, most of the information 
concerning HD is expected to lie above 1000 \AA, in the LiF channels.

The spectra were extracted from the 11 sub-exposures with the version 1.6 of 
the pipeline, and the electronically induced detector bursts were subtracted 
by screening of the events with IDL tools. The processing includes data 
screening, thermal drift correction, geometric distortion correction, Doppler 
correction to heliocentric wavelengths, dead time correction (data loss due to 
overflow of photon events) and wavelength calibration, but not correction for 
astigmatism. In order to obtain the best resolution, the calibrated spectra 
were then manually coadded over small wavelength windows, using both exposure 
time and error weighting. The resulting signal to noise ratio is 
$\simeq20$~per resolution element in the continuum near the strongest 
available HD lines.

Fig. 1 shows the final recorded spectrum. Broad H$_2$~bands are clearly seen 
(see also Snow et al. 2000), together with some CO bands and several other 
narrow interstellar features including atomic species such as C{\sc i}, 
O{\sc i}, N{\sc i}, Fe{\sc ii} and P{\sc ii}. One can note that the heavily 
saturated H$_2$~lines reach a zero flux, indicating an extremely low level of 
scattered light in the {\it FUSE} instrument.

We have performed a preliminary analysis of these data by using a simultaneous 
profile fitting of all detected lines from any molecular and atomic species 
included in a recent updated compilation by Morton (private communication). 
In a first step, a single absorbing component was assumed (and found at a
coherent radial velocity for all lines). It has to be noted that most of the 
lines are saturated so that the column densities are highly sensitive to the 
value of the intrinsic broadening parameter -- $b$-value -- which combines 
thermal and turbulent widening. Furthermore, the uncertainties in the 
instrumental line-spread function make it quite plausible that large amounts 
of cold absorbing gas might be "hidden". Nevertheless, the dominant $J=0$~and 
$J=1$~levels of the H$_2$~lines lie on the damped part of the curve of growth 
and the total H$_2$~column density is almost independent of the assumed 
$b$-value and the velocity structure of the line of sight. We find 
$\log_{10}N({\rm H}_2)_{tot}$ = 21.2 $\pm 0.2$, in agreement with Snow et al. 
(2000).

Absorption from R(0) lines of HD are also detected toward HD 73882, four 
examples of which are presented in Fig. 2, arbitrarily set to 
0 km s$^{-1}$~radial velocity. This is a $\sim5\sigma$~detection, as shown in 
the upper panel of Fig. 3, in which is plotted the $\Delta\chi^2$~of the 
simultaneous fits of all detected HD lines (except the one blended with 
O{\sc vi} near 1031.9 \AA) as a function of the HD column density: when 
$N({\rm HD})$~decreases, $\Delta\chi^2$~tends to nearly 25. The HD lines, as 
well as the CO and the high $J$~level H$_2$~ones, lie on the flat part of the
curve of growth where $N({\rm HD})$~heavily depends upon the assumed effective 
$b$-value (note that this makes useless the use of equivalent widths). In that
case, the assumption of a single absorbing component is indeed critical for 
deriving column densities. From the upper panel of Fig. 3, allowing, 
conservatively, for 3$\sigma$~variations (about 10 for $\Delta\chi^2$), we 
find a huge range of values 14.6 $< \log_{10}N({\rm HD}) <$ 16.9, the lower 
end being even larger than most of the {\it Copernicus} measurements. From the 
present data alone, the saturation of the lines prevents a more precise 
determination. However, one sees that the best estimates 
($\Delta\chi^2 \sim 0$) are found for $\log_{10}N({\rm HD})$~around 16.1 
(solid line in Fig. 2) which corresponds, according to the bottom panel of 
Fig. 3, to an effective $b$-value of about 1 km s$^{-1}$.

%\begin{figure*}
%\plotone{chi2.ps}
%\caption[]{The upper panel is a plot of the $\Delta\chi^2$~of the spectral 
%           fits as a function of the HD column density. The lower panel gives 
%           the best estimates of $N({\rm HD})$~as a function of the effective 
%           $b$-value. Small $b$-values are slightly favoured by the 
%           $\Delta\chi^2$.}
%\label{chi2}
%\end{figure*}

\section{Discussion.}

Previous observations of molecules toward HD 73882 include CH, CH$^+$, CN,
C$_2$ and CO (Gredel, van Dishoeck \& Black 1993, 1994). Although the 
molecular fraction 2 $N({\rm H_2})/[N(H) + 2 N({\rm H_2})]$~is quite close to
that found toward $\zeta$~Oph (Snow et al. 2000), the CO/H$_2$~and 
CN/H$_2$~ratios, being very sensitive to the radiation field, differ greatly.
These ratios are over 5 times larger for HD 73882 and closer to values found 
for other dark clouds such as TMC1. The extinction curve for HD 73882 is 
peculiar compared to "normal" diffuse cloud curves, with a steep far-UV rise 
and a weak bump at 2200 \AA. Moreover, the value of 
R$_{\rm v}$=A$_{\rm v}$/$E_{\rm B-V}$~is 3.39, larger than the mean galactic 
value $\simeq3$~(Cardelli et al. 1989).

Ground based mm-wave $^{12}$CO reveal three components in {\it emission} 
separated by 3 and 2.3 km s$^{-1}$~(i.e. 5.3 km s$^{-1}$~for the extreme ones,
nearly as for $^{13}$CO; Gredel, van Dishoeck \& Black 1994), not resolved 
with {\it FUSE} spectra if indeed they are really intercepted by the HD 73882 
line of sight. Snow et al. (2000) show that Na{\sc i} has many components. 
However, all diatomic molecules seen in {\it absorption} show only one 
component, including the C$_2$~observations recorded at high spectral 
resolution (R=10$^5$). These components agree in velocity. Therefore, several 
components might be present, but the possibility of a single molecular 
absorbing region on the HD 73882 line of sight is not excluded.

The present observation of this relatively high extinction target opens the
possibility that, indeed, we are probing a cloud which could have reached the
transition point where the reservoir of deuterium is primarily the HD molecule.
We have modeled the photodissociation region with a density of 350~cm$^{-3}$,
as determined from the analysis of C$_2$ excitation (Gredel, van Dishoeck \& 
Black 1993), a standard ultraviolet interstellar radiation field and a cosmic 
ionization rate $\xi$ = 5 $\times$ 10$^{-17}$ s$^{-1}$, considered as an 
average (Abgrall et al. 1992; Le Bourlot et al. 1993; Roueff \& 
Nod\'e-Langlois 1998). The atomic-to-molecular transition is assumed to be
represented by a semi-infinite, plane parallel slab with the radiation field 
impinging on one side only of the cloud. The mechanisms involved in the 
photodissociation of HD are very similar to those for H$_2$, but self-shielding
is much more efficient for the abundant H$_2$~than for HD. Similarly to 
H$_2$, HD is formed on grains. However, HD is also predominantly formed 
through the reaction $H_{2} + D^{+} \rightarrow HD + H^{+}$, whenever 
substantial H$_2$~is present.

The main results of this model are displayed in Fig. 4 which shows the 
evolution of the fractional abundances of atomic hydrogen and deuterium, and 
their associated molecular species, as a function of the extinction 
A$_{\rm v}$. It is clear that the H to H$_2$~transition occurs at 
significantly lower extinction than the D to HD transition (note that the C to 
CO transition would take place at even higher extinction than HD). This
reflects the smaller contribution of self-shielding in the photodissociation 
mechanism for HD. According to the extinction value for our target star, if 
there are not a large number of velocity components, we might be dealing with 
an absorbing region where HD is the primary reservoir of deuterium. It also 
shows that even if several components are detected in atomic lines, they may 
be present in the H$_2$~lines, but not necessarily in the HD ones which are
tracing a deeper region in the cloud.

\section{Conclusion.}

We have reported {\it FUSE} observations in the direction of the reddened 
star HD 73882. Several lines from the HD molecule are detected, together
with many from H$_2$. Under the assumption of a single absorbing molecular 
component, our best estimate for the HD/H$_2$~column density ratio from a 
simultaneous profile fitting of all detected lines is of the order of 
$10^{-5}$. However, strictly speaking this is only an upper limit since the 
fit is very sensitive to the intrinsic width of the lines, most of them being 
saturated.

This preliminary evaluation toward such a reddened star opens the possibility 
of probing regions sufficiently deep into interstellar molecular clouds for HD 
to be the main reservoir of deuterium atoms. In such regions, the measurement 
of the HD/H$_2$~ratio would thus become a new reliable and powerful tool for 
determining the interstellar D/H ratio in dense regions for which the simple 
relation \dsh~=~0.5 $\times~$HD/H$_2$~will apply. Recall that the possibility
of multiple clouds for the present line of sight prevents claiming it is 
indeed the case. One may hope that a number of similar future {\it FUSE} 
observations will ultimately succeed, providing additional information on the 
structure of the observed sight-lines is available.

\
%-------------------- acknowledgments -----------------
\acknowledgments
%------------------------------------------------------
{\bf Acknowledgments.} This work is based on data obtained for the Guaranteed 
Time Team by the NASA-CNES-CSA FUSE mission operated by the Johns Hopkins 
University. Financial support to U.S. participants has been provided by NASA
contract NAS5-32985. We thank the anonymous referee for his careful reading
of the manuscript.

%{\it FUSE} is an {\it Origins} Mission. It is funded 
%by NASA Explorer Program in cooperation with the Canadian Space Agency and 
%the Centre National d'\'Etudes Spatiales of France. {\it FUSE} was developed 
%and is being operated for NASA by the Johns Hopkins University in 
%collaboration with the University of California, Berkeley and the University 
%of Colorado. Financial support has been provided by NASA contract NAS5-32985.

\clearpage

\figcaption[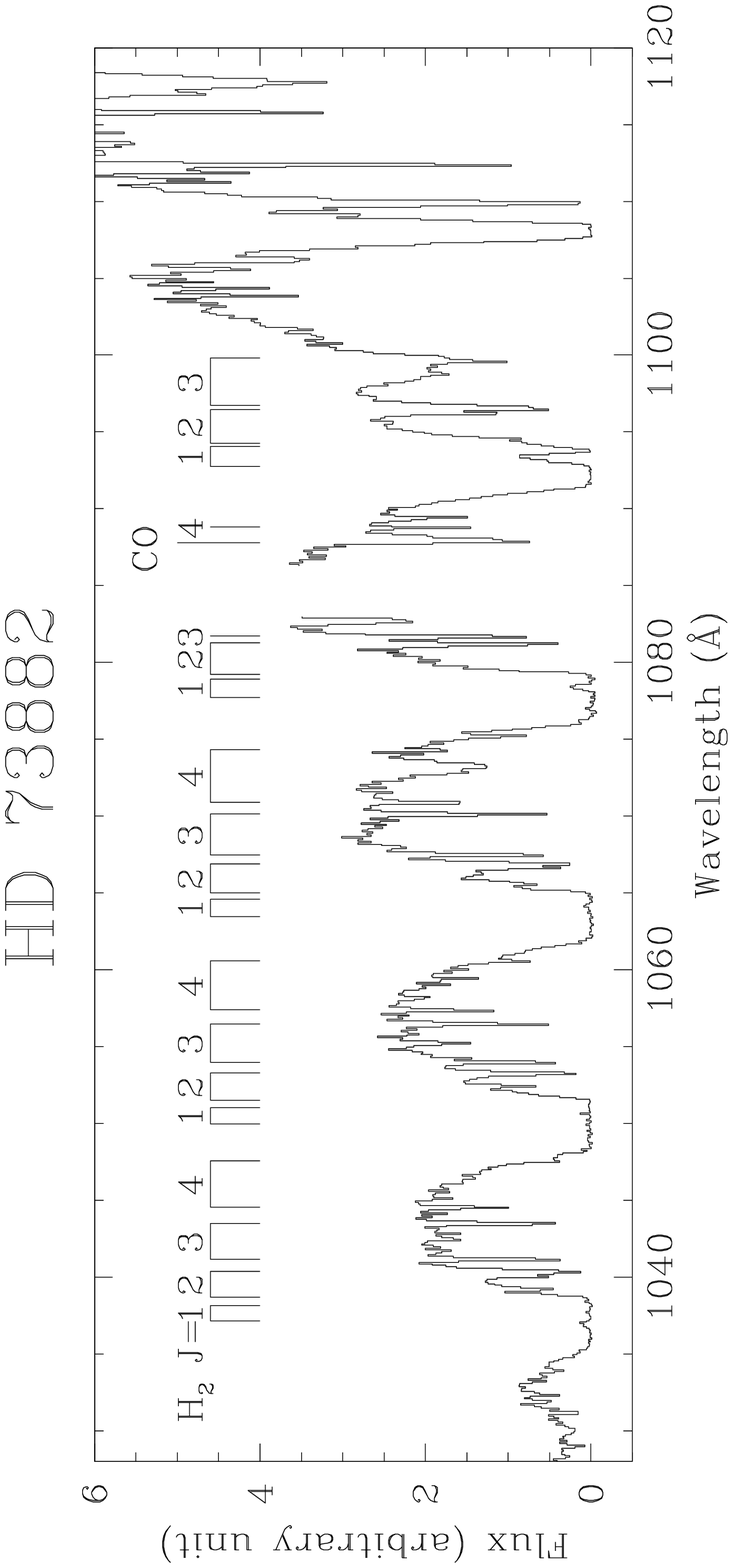]{{\it FUSE} spectrum of HD 73882 from 1030 to 
1120 \AA. Many atomic and molecular absorption lines are detected. Most of 
them are from molecular hydrogen (tick marks indicate those from $J=1$~to 
$J=4$). Lines from H$_{2}(J=0)$~at 1037, 1049, 1063, 1077, 1092 and 1108 
\AA~and from H$_{2}(J=1)$~are strongly saturated. A CO band at 1088 \AA~is 
also indicated. The gap around 1085 \AA~is due to the gap in between the two 
micro-channel plates mounted in each LiF detectors. \label{fig1}}

\figcaption[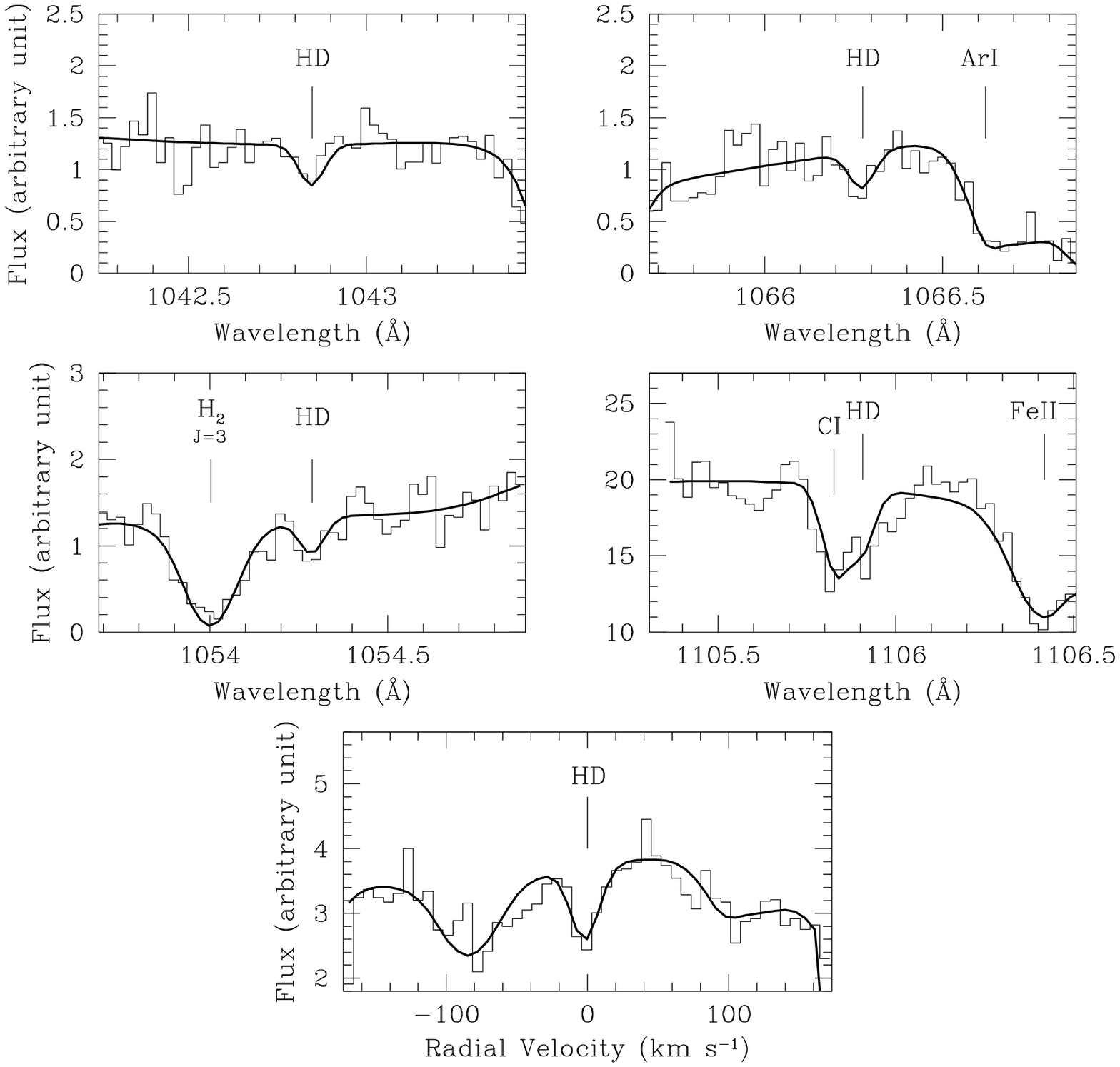]{Plots of four 
$B^{1}\Sigma^{+}_{u}$--$X^{1}\Sigma^{+}_{g}$~electronic transitions of HD 
detected in absorption toward HD~73882 at 1042.85, 1054.29, 1066.27 and 
1105.91 \AA, corresponding respectively to J=5$\rightarrow$0, 
J=4$\rightarrow$0, J=3$\rightarrow$0 and J=0$\rightarrow$0 (the 
J=2$\rightarrow$0 and J=1$\rightarrow$0 lines are hidden by strong 
H$_2$~lines). The actual {\it FUSE} wavelength absolute calibration being 
uncertain by few km s$^{-1}$, the wavelength scale has been shifted and the 
HD lines aligned at the same radial velocity arbitrarily set to 0 km s$^{-1}$. 
The solid line is the best fit obtained simultaneously for all detected HD 
lines and other identified absorptions. To underline the detection 
independently proven by the $\chi^2$~variation (see text and Fig.~\ref{chi2}), 
the bottom panel shows the composite solution covering the first three HD 
lines, thus enhancing the S/N ratio (the line at 1105.91 \AA~is blended with 
C{\sc i} and has been excluded in the summation). \label{fig2}}

\figcaption[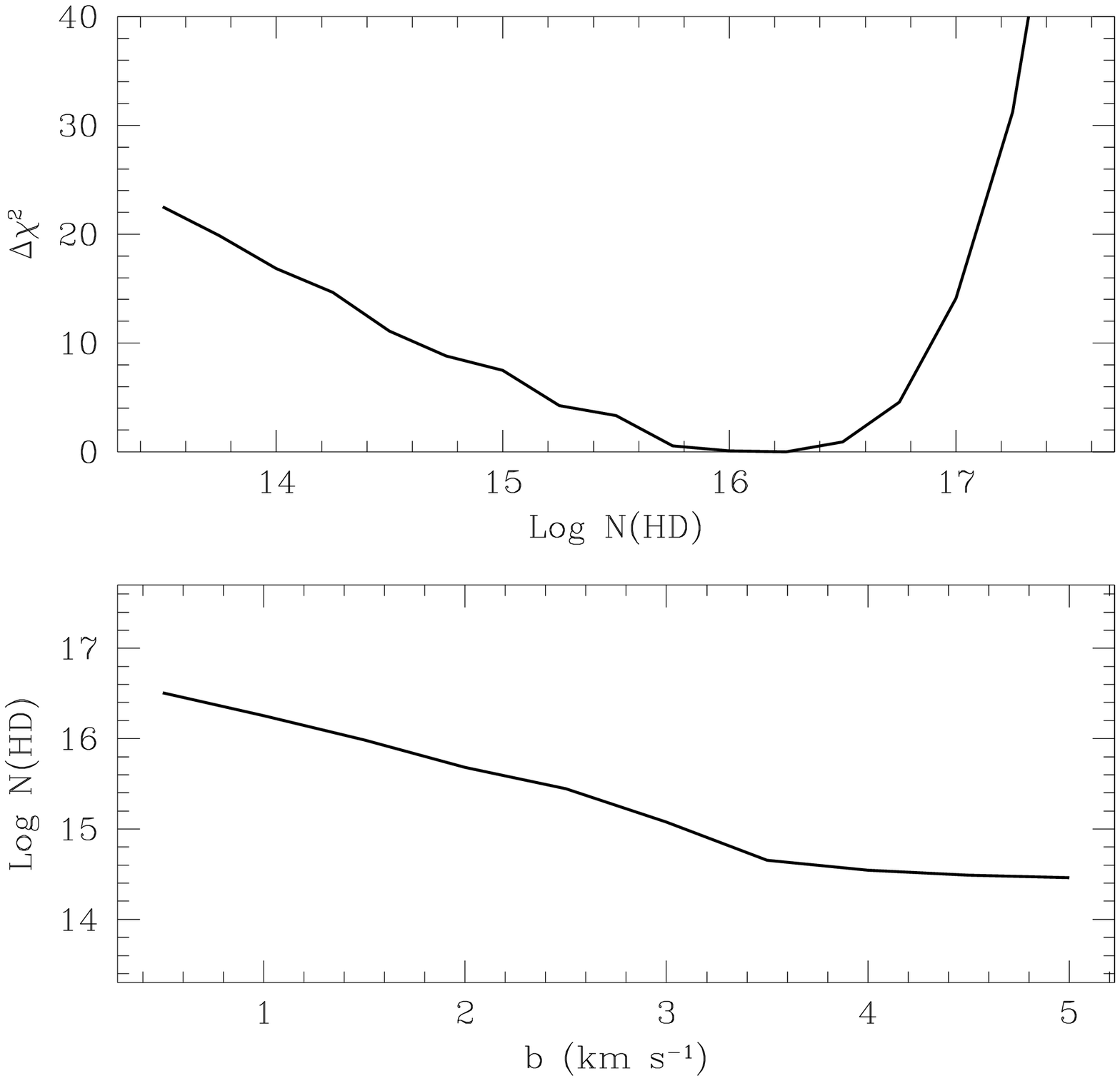]{The upper panel is a plot of the $\Delta\chi^2$~of the 
spectral fits as a function of the HD column density. The lower panel gives 
the best estimates of $N({\rm HD})$~as a function of the effective $b$-value. 
Small $b$-values are slightly favoured by the $\Delta\chi^2$. \label{fig3}}

\figcaption[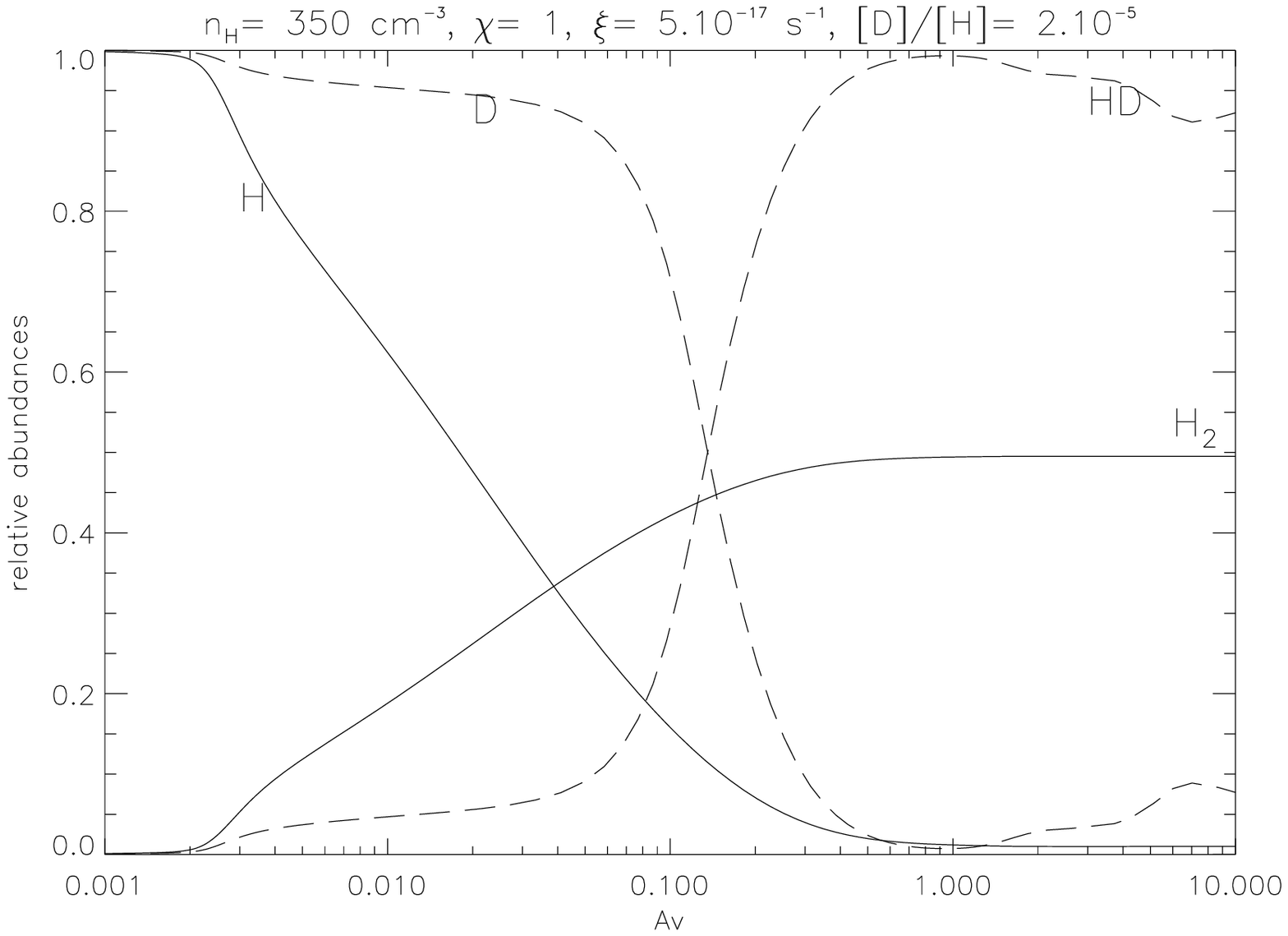]{Comparison of the relative evolution of H, D, 
H$_{2}$~and HD fractional abundances (relative to the proton density) with 
A$_{\rm v}$. We assume that a semi infinite plane-parallel cloud is exposed to
an isotropic "standard" interstellar ultraviolet radiation field (scaling 
factor $\chi$=1). The cloud has a constant density $n_{\rm H}$=350~cm$^{-3}$, 
deduced from C$_2$ observations toward HD 73882. $\xi$ is the cosmic 
ionization rate. For HD 73882, A$_{\rm v}$=2.44. The small changes in the HD 
and D curves beyond A$_{\rm v}\sim$5 are due to slight variations in 
temperature within the cloud which in turn induce slight chemical variations.
\label{fig4}}

\end{document}